\documentclass[letter,scriptaddress,twocolumn, showkeys]{revtex4}
\usepackage{ amssymb }
	\usepackage{amsmath}%,amssymb} 
	\usepackage{makeidx}
	\usepackage{amsfonts}
	\usepackage[ansinew]{inputenc}
	\usepackage[usenames,dvipsnames]{pstricks}
	\usepackage{subfigure}
	\usepackage{epsfig}
	\usepackage{pst-grad} % For gradients
	\usepackage{pst-plot} % For axes
		\usepackage{amsthm}
\usepackage{ amssymb }
 	\usepackage{makeidx}
	\usepackage{amsfonts}
 	\usepackage{listings}
 	\usepackage{mathtools}
	\usepackage{dblfloatfix}
	\usepackage{float}
	
	\usepackage[colorlinks,hyperindex]{hyperref}
	\hypersetup
	{
		colorlinks,%
		citecolor=blue,%
		linkcolor=blue,%
		urlcolor=blue,%
	}
	\usepackage{tikz}

\usepackage{amsmath}
\usepackage{amssymb}
\usepackage{graphicx}
\usepackage{xcolor}
\usepackage{float,graphicx}
\usepackage{placeins}
\usepackage{color, colortbl}
\usepackage[colorinlistoftodos]{todonotes}
\theoremstyle{definition}

	\theoremstyle{plain}

\makeindex

\begin{document}

\title{A modified age-structured SIR model for COVID-19 type viruses}

\author{Vishaal Ram$^{a}$, Laura P. Schaposnik$^{\star, b}$
}
% \affiliation[label1]{}
  \affiliation{($\star$) Corresponding author: schapos@uic.edu}

\begin{abstract}
{ \footnotesize Abstract:
We present a modified age-structured SIR model based on known patterns of social contact and distancing measures within Washington, USA. We find that population age-distribution has a significant effect on disease spread and mortality rate, and contribute to the efficacy of age-specific contact and treatment measures. We consider the effect of relaxing restrictions across less vulnerable age-brackets, comparing results across selected groups of varying population parameters. Moreover, we analyze the mitigating effects of vaccinations and examine the effectiveness of age-targeted distributions. Lastly, we explore how our model can applied to other states to reflect social-distancing policy based on different parameters and metrics. }
 \end{abstract}

\maketitle 

\section{Introduction}

The study of the spread of diseases and rumours within networks (social and biological) in order to  trace factors that are responsible for or contribute to their occurrence has been done from many different perspectives. Moreover, only recently have graph theory, number theory, and computer science taken researchers to several breakthroughs. Back in the early 1900s,  Ronald Ross produced the first mathematical model of mosquito-borne pathogen transmission using mosquito spatial movement in order to reduce malaria from an area \cite{Ross}. Some decades later, William O. Kermack and Anderson G. McKendrick \cite{Kermack} created the \textit{SIR model}, which categorized people into the $3$ states {\it Susceptible}, {\it Infectious} and {\it Removed} -- the model which we shall focus on. 

More recently,  contact networks were introduced to better represent a community \cite{Keeling}: these are   adapted to reflect certain particular characteristics of society, and they have been of much use when doing mathematical modeling of epidemics. In this setting, a social network is modelled as a graph where vertices represent individuals, and edges encode the interactions amongst people: two people are connected by an edge in the graph whenever they are related (and thus an interaction could exist).

Given the recent outbreak of COVID-19, and with views towards applications to future viral outbreaks and marketing strategies, this paper is dedicated to the study of contention strategies with social networks by targeting different clusters within the network in different ways.   As highlighted in \cite{Pastor}, the importance of local clustering in networks has been widely recognised, and not much study has been done in this direction until very recently.

Since  evidence shows very large differences in hospitalization and fatality rates between age groups and gender groups, our interest is on obtaining a modified age-structured SIR model.  
Very recently, a first step in analyzing the role of optimal targeted lockdowns in
a multi-group extension of the standard SIR model was done \cite{age1,acemoglu2020optimal}, where it was found that  among strategies which end with population immunity, strict age-targeted mitigation ones have the potential to greatly reduce mortalities and ICU utilization for natural parameter choices \cite{age1}. Moreover, the trade-offs facing policy-makers between saving lives and improving economic outcomes were analized in  \cite{acemoglu2020optimal}, where it is shown that better social outcomes are possible with targeted policies:  ``{\it  Differential lockdowns on groups with differential risks can significantly improve policy trade-offs, enabling large reductions in economic damages or excess deaths
or both''} \cite{acemoglu2020optimal}.

In the present work we take different path from \cite{age1,acemoglu2020optimal} and consider an age-compartment model  with a rescaling function completely based on the policy that Washington implements, where the intensity of the social distancing policy is proportional to the ICU occupancy. It should be noted that a modified rescaling could be applied to other states, hence making our model adaptable to other settings, e.g.~New York uses metrics including rate of change of total infections in their policy. Moreover, we consider age-specific relaxation policy (e.g.~opening schools/work) and vaccine distribution. By applying our   model  to populations of varied age-distribution, we see the following:
\begin{itemize}
 \item  Following our rescaling function, population age-distribution is directly correlated with increasing peak ICU occupancy and decreasing peak infection count. However, herd immunity threshold is unaffected by the change in population parameters with the same proportion of the population being infected through the course of the epidemic. 
\item Across all age-distributions, relaxing school and work restrictions has the effect of infecting the same proportion of the population across a smaller time frame, increasing peak ICU occupancy by over to $18 \%$ and $51 \%$ respectively. However, such effects are not observed when relaxing restrictions after the initial peak in infections. 

\item Administering vaccines at a constant rate lowered the herd immunity threshold, especially among high median age counties, while also reducing mortality rate by $28 \%$. Moreover, strictly prioritizing vaccines to older age-brackets seems extremely effective, lowering ICU occupancy and further reducing mortality rate by $20 \%$ while also completely preventing the spread of the virus in the short term.
\end{itemize}

To illustrate our perspective, we study the available data from the state of Washington, USA, and apply our modified model to this dataset. Our paper is organized as follows: we shall begin by introducing the SIR model  in Section \ref{SIR1}, and an age structured version following the work in \cite{age1} in Section \ref{SIR2}.

\newpage
\section{The (Age-Structured) SIR Model} \label{SIR}
As mentioned previously, the SIR model is a simple model for infectious disease in which the population is divided into three compartments: those susceptible to the disease, those infected with the disease, and those removed from the disease either through death or recovery. Across this paper, we shall assume that those in the removed group are unable to be infected again.

\subsection{The SIR model}\label{SIR1}
The number of individuals in each group is given by certain functions of time $S(t)$, $I(t)$, $R(t)$ respectively. Moreover, the dynamics of the model are given by the set of ordinary differential equations:

\begin{eqnarray}
    \frac{\mathrm dS}{\mathrm dt} &=& -\beta \cdot I \cdot \frac{S}{N}; \\
    \frac{\mathrm dI}{\mathrm dt} &=& \beta \cdot I \cdot \frac{S}{N} - \gamma \cdot I; \\ 
    \frac{\mathrm dR}{\mathrm dt} &=& \gamma \cdot I,
\end{eqnarray}
 which depend on the following parameters:
\begin{itemize}
    \item the total population $N$;
    \item  the transmission rate $\beta$, measured as the average number of contacts per person per time, multiplied by the probability of transmission between a infected and susceptible person;
    \item  and the removal rate $\gamma$, also given by $1/D$ where $D$ is the length of the period for which a person is infectious.  
\end{itemize}

During the early stages in an epidemic, transmissions between individuals are statistically independent, meaning that the probability that an infectious individual encounters someone no longer susceptible is probabilistically low. Within the model, the \textit{basic reproduction number} $R_0$ is he number of people an individual is expected to infect,  and  can be computed given the parameters of the SIR model as $R_0 = \frac{\beta}{\gamma}$. 

 One should note that the $R_0$ value is not a biological constant as its value depends on factors such as individual contact patterns. However, the number $R_0$ of a disease is generally consistent among newly susceptible populations and can be used to predict the trajectory of an epidemic or calibrate the initial conditions of a model. In particular, a value of $R_0>1$ indicates a disease will begin to spread in a population if no contention is installed, where a greater $R_0$ value indicates faster exponential growth. For example,  {\it measles} is known to be   one of the most contagious diseases, with $12\leq R_0\leq18$, which means that each measles-infected person may spread the virus to 12 to 18 other individuals in a susceptible population \cite{guerra2017basic}.
For comparison, the CDC estimates that COVID-19 has an $R_0$ value of about $5.7$ in the United States \cite{CDCR0}, close to that of Polio and Rubella.

\subsection{An Age-Structured SIR Model}\label{SIR2}
For many diseases such as COVID-19, the effect on different age-groups varies drastically. Therefore, we consider an age structured model in which we compute the age distribution of each compartment in each of the age-brackets 0-9, 10-19, \dots, 70-79, and 80+. This separation, in particular, is much more specific than the one used in \cite{acemoglu2020optimal} and thus allows us to have our results in a more refined way.  

For an age-structured model, we must incorporate an age-contact matrix $\mathcal{M}$ describing the rate of contact between each pair of age-brackets. In the present paper, we shall use the same matrix used in \cite{age1} based on data collected by \cite{Age2} for the United States,   shown in Figure \ref{fig:cmat1} below. In this setting, the values in $\mathcal{M}$ are proportional to the total number of contacts per time between age-brackets, divided by the product of their population sizes. In particular, $\mathcal{M}$ would be a constant matrix if individuals were equally likely to contact each other across all age-brackets.

\begin{figure}[H]
    \begin{center}
    \includegraphics[scale = 0.65]{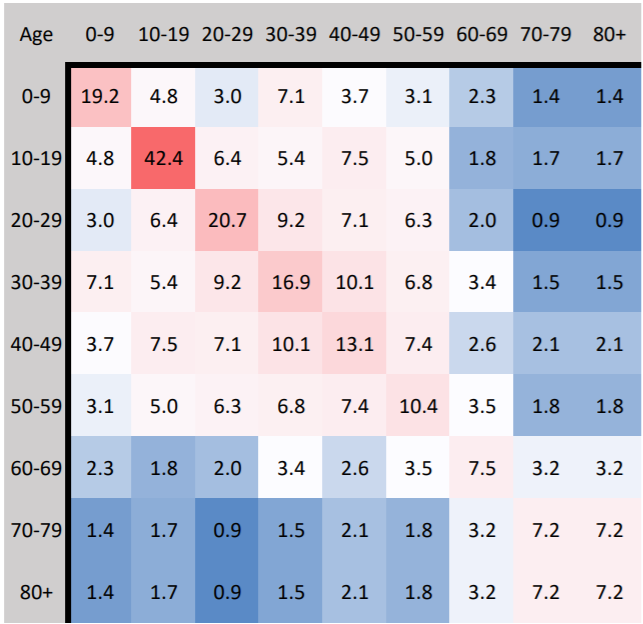}
    \end{center}
    \caption{Age-Contact Matrix $\mathcal{M}$.}
    \label{fig:cmat1}
\end{figure}

Following \cite{age1}, in our age-structured SIR model we define vector valued functions $S(t)$, $I(t)$, and $R(t)$ representing the age-distribution of the total individuals susceptible, infectious, and removed respectively by letting the $ith$ coordinate indicate the number of individuals  in the $i$th age-bracket  for $1 \le i \le 9$. Then, from \cite{age1}, the dynamics of the model are given by the following equations:
\begin{eqnarray}
    \frac{\mathrm dS_i}{\mathrm dt} &=& -\beta \cdot \frac{S_i}{N} \cdot \sum_{j=1}^n \mathcal M_{ij} \cdot I_j \\ 
    \frac{\mathrm dI_i}{\mathrm dt} &=& \beta \cdot \frac{S_i}{N} \cdot \sum_{j=1}^n \mathcal M_{ij} \cdot I_j - \gamma \cdot I_i \\ 
    \frac{\mathrm dR_i}{\mathrm dt} &=& \gamma \cdot I_i 
\end{eqnarray}
 
 Let vector $\mathbf{p}$ denote the proportion of the population in each age-group, and let $\lambda$ and $\mathbf{v}$ be the dominant eigenvalue and corresponding eigenvector of $ \mathcal{M} \cdot  \text{diag}(\mathbf{p})$. In the initial state of the epidemic, the growth rate of transmissions follows a steady state, i.e. $I \propto \frac{\mathrm dI}{\mathrm dt}$. It is shown in \cite{age1} that in this state, the value of $R_0$ can be computed as $\frac{\beta \cdot \lambda}{\gamma}$, with the initial infected distributed according to $\mathbf{v}$. Therefore to emulate the $R_0$ value of COVID-19, we can assign $\beta = \frac{R_0 \cdot \gamma}{\lambda}$ where $\gamma = \frac{1}{14}$, indicating a 14-day infectious period. 

Consider the vectors $\mathbf{h} = \{h_1,\dots,h_9 \}$, $\mathbf{c} = \{c_1,\dots,c_9 \}$, and $\mathbf{m} = \{m_1,\dots,m_9 \}$ to be the hospitalization rate, ICU rate among hospitalizations, and mortality rates, respectively for each age-bracket labeled by $i$. Then, one can  compute the vector valued functions $H(t)$, $C(t)$,  and $M(t)$ representing the age-distribution of the total individuals hospitalized, in critical care, and deceased respectively through the following differential equations:
\begin{eqnarray}
    \frac{\mathrm dH_i}{\mathrm dt}  &=& \gamma \cdot h_i \cdot I_i \\ 
    \frac{\mathrm dC_i}{\mathrm dt} &=& \gamma \cdot h_i \cdot c_i \cdot I_i \\ 
    \frac{\mathrm dM_i}{\mathrm dt} &=& \gamma \cdot m_i \cdot I_i  
\end{eqnarray}

\begin{figure}[H]
    \begin{center}
    \includegraphics[scale = 0.8]{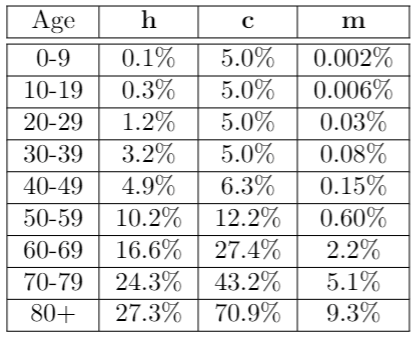}
    \end{center}
    \caption{COVID-19 Age Statistics}
    \label{fig:agetable}
\end{figure}

In what follows we shall use the COVID-19 estimates for these values from \cite{stats} shown in Figure \ref{fig:agetable} to understand the above functions. 

\section{Social Distancing in Washington State, USA.}

In what remains of the manuscript, we shall pay special attention to the COVID-19 outbreak that took place in Washington State, USA since January 2020, and use the data available to model different Social Distancing strategies.  The first confirmed case of the COVID-19 pandemic in the United States was announced in Washington State on January 21, 2020. Five weeks later, on February 29th, Washington also announced the first COVID-19 related death in the country. On March 23, Governor Jay Inslee issued the first stay-at-home order which lasted until the end of May \cite{online}. 

On May 29th, Inslee announced a \href{https://www.governor.wa.gov/sites/default/files/SafeStartPhasedReopening.pdf}{Safe Start}: a four phased county-by-county reopening plan. The plan allows counties to gradually relax social-distancing measures based on their assessments of health care system readiness, testing capacity and availability, case and contact investigations, and ability to protect high-risk populations. One of the main factors determining a county's reopening procedure is the percentage of ICU beds available in hospitals. Therefore to model the effect of social-distancing policy it is useful to scale the contact matrix $\mathcal{M}$ by a value proportional to this percentage:  
\begin{eqnarray} \mathcal{M} \to \frac{1}{\lambda \cdot |C|/C_{\text{max}}} \mathcal{M}\end{eqnarray}
where $|C|$ is the total number of individuals in critical care while $C_{\text{max}}$ is the ICU capacity which we set to the US average of 34.7 per 100,000 residents. As we mentioned before,  this leads to our age-compartment model, utilizing a rescaling function completely based on the policy that Washington implements where the intensity of the social distancing policy is proportional to the ICU occupancy. We refer to $\frac{1}{\lambda \cdot |C|/C_{\text{max}}}$ as the mitigation factor. The constant $\lambda$ determines the ``strictness'' of the social-distancing measures. Such reactive mitigation measures have been done before in SIR models (e.g., see \cite{react} with respect to total infected count). 

\begin{figure}[H]
    \begin{center}
    \includegraphics[scale = 0.14]{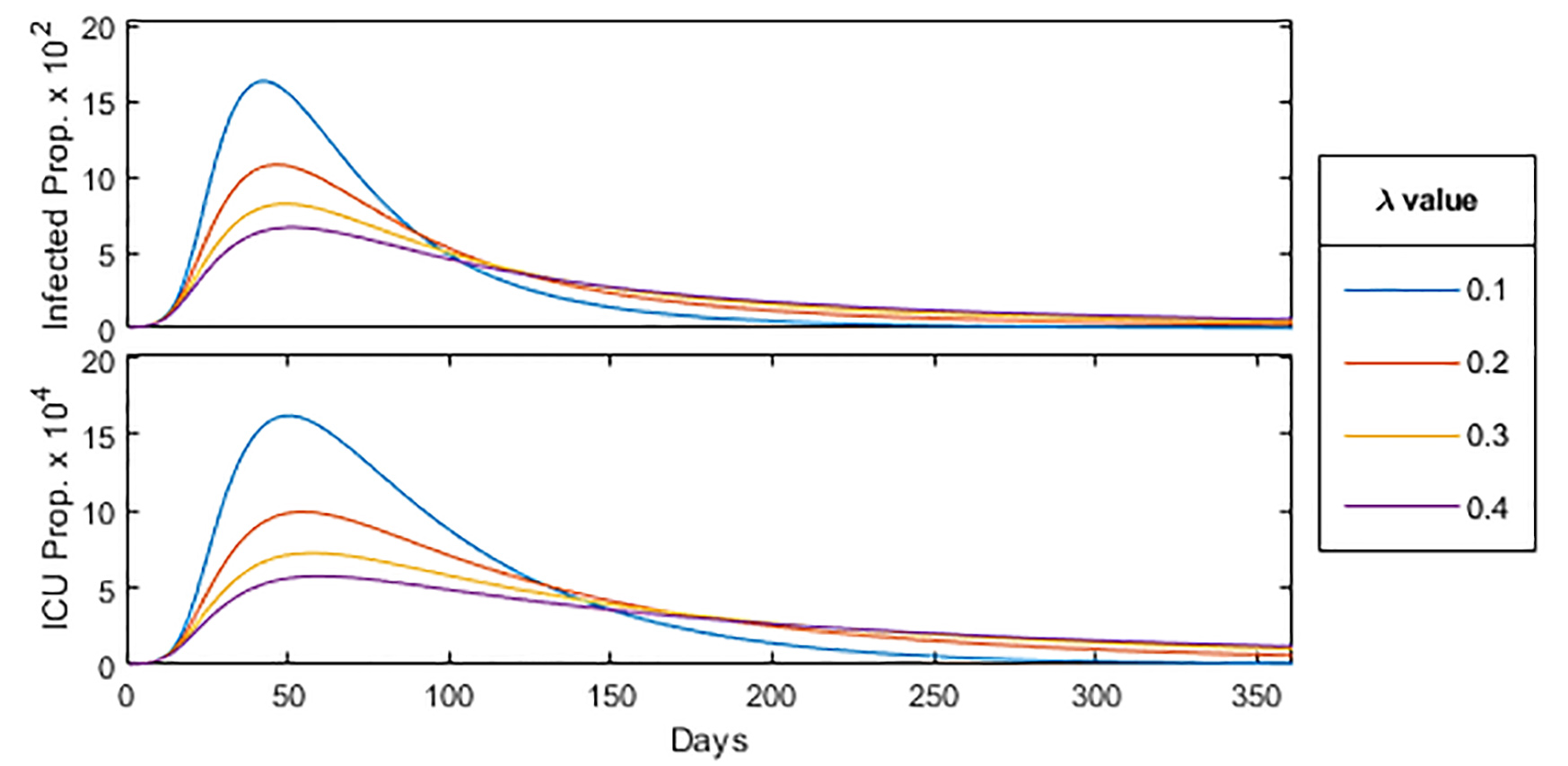}
    \end{center}
    \caption{Effect of $\lambda$ in SIR-model. 
}
    \label{fig:lambda}
\end{figure}

In order to understand the implications of the constant $\lambda$, we should note that a larger $\lambda$ value has the effect of ``flattening the curve'', decreasing total case count while also slowing the rate of decline in cases.   Moreover, one can see in Figure \ref{fig:lambda}  the effect of $\lambda$ on the proportion of the population (with Washington state demographic parameters) infected and in the ICU ($I$ and $C$ respectively). For consistency, we use a $\lambda$ value of $0.1$ in our modeling. 

\subsection{Demographic Parameters}
\begin{figure}[H]
    \begin{center}
    \includegraphics[scale = 0.5]{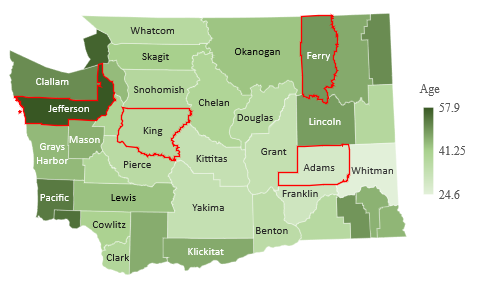}
    \end{center}
    \caption{Median Age.}
    \label{fig:med}
\end{figure}

\begin{figure}[H]
    \begin{center}
    \includegraphics[scale = 0.5]{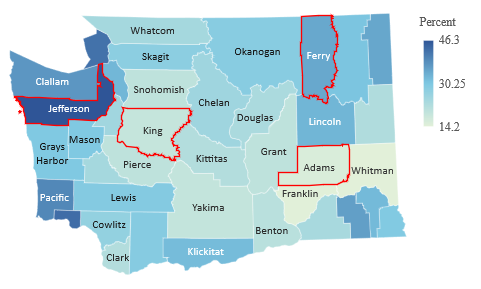}
    \end{center}
    \caption{Percent of population over 60.}
    \label{fig:per60}
\end{figure}

\begin{figure}[H]
    \begin{center}
    \includegraphics[scale = 0.5]{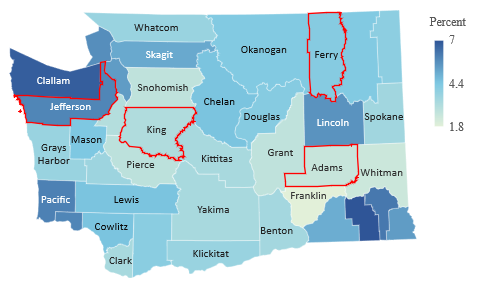}
    \end{center}
    \caption{Percent of population over 80.}
    \label{fig:per80}
\end{figure}

The main objective of our model is to investigate the effect of a population's age distribution on the transmission and spread of a virus like COVID-19, which is both highly contagious and largely age-specific in its effect on the population. To aid in our comparison of populations, we select four sample counties of Washington state with varying age-distribution: Jefferson, King, Ferry and Adams (outlined in red in Figures \ref{fig:med}, \ref{fig:per60}, \ref{fig:per80}) and apply \textcolor{blue}{our model} to their demographic parameters. 

For each county, we use the same parameters specific to COVID-19, but adjust the initial state of the function $S(t)$ based on the age-distribution of the county, which can be seen in Figure \ref{fig:med}, obtained via the official government's data in \cite{online2}. 

As shown in Figure \ref{fig:med}, there is significant variation in the age distribution of each county with counties along the west coast such as Jefferson and Clallam with a median age of over 50 years, while counties in central and eastern Washington such as Adams and Whitman with a median age of under 30 years. Due to this variation, we expect  an epidemic such as  that of COVID-19 to have similar variation in its effect on the population, from rate of spread to mortality rate, and thus warrant different, age-targeted mitigation measures. 

In order to understand the the relevance of the mortality rate within our study, there are two other important data points to consider, which  are the proportion of the population in each county over the age of 60 and 80, shown in Figures \ref{fig:per60} and \ref{fig:per80} respectively. These two population groups represent the age-brackets most vulnerable and are generally a better indicator of the overall population mortality rate. 

\section{Model Application}

\begin{figure}[H]
    \begin{center}
    \includegraphics[scale = 0.5]{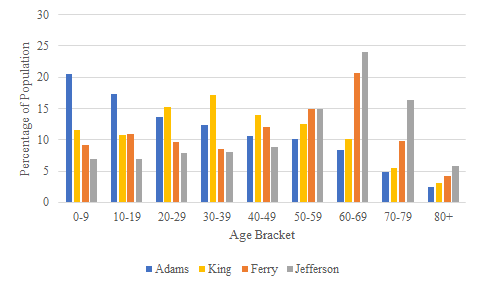}
    \end{center}
    \caption{Age Distribution of Selected Counties.}
    \label{fig:ageDist}
\end{figure}

As mentioned previously, the four counties of Washington we selected for our model comparison are Adams, King, Ferry, and Jefferson. These counties have   median ages of 28.3, 36.8, 49.2, and 57.9 respectively. The complete age distribution of the selected counties are shown in Figure \ref{fig:ageDist}, and in this section we shall apply our SIR model on each of the four selected counties. 

\begin{center}

 \begin{figure}[H]
    \includegraphics[scale = 0.15]{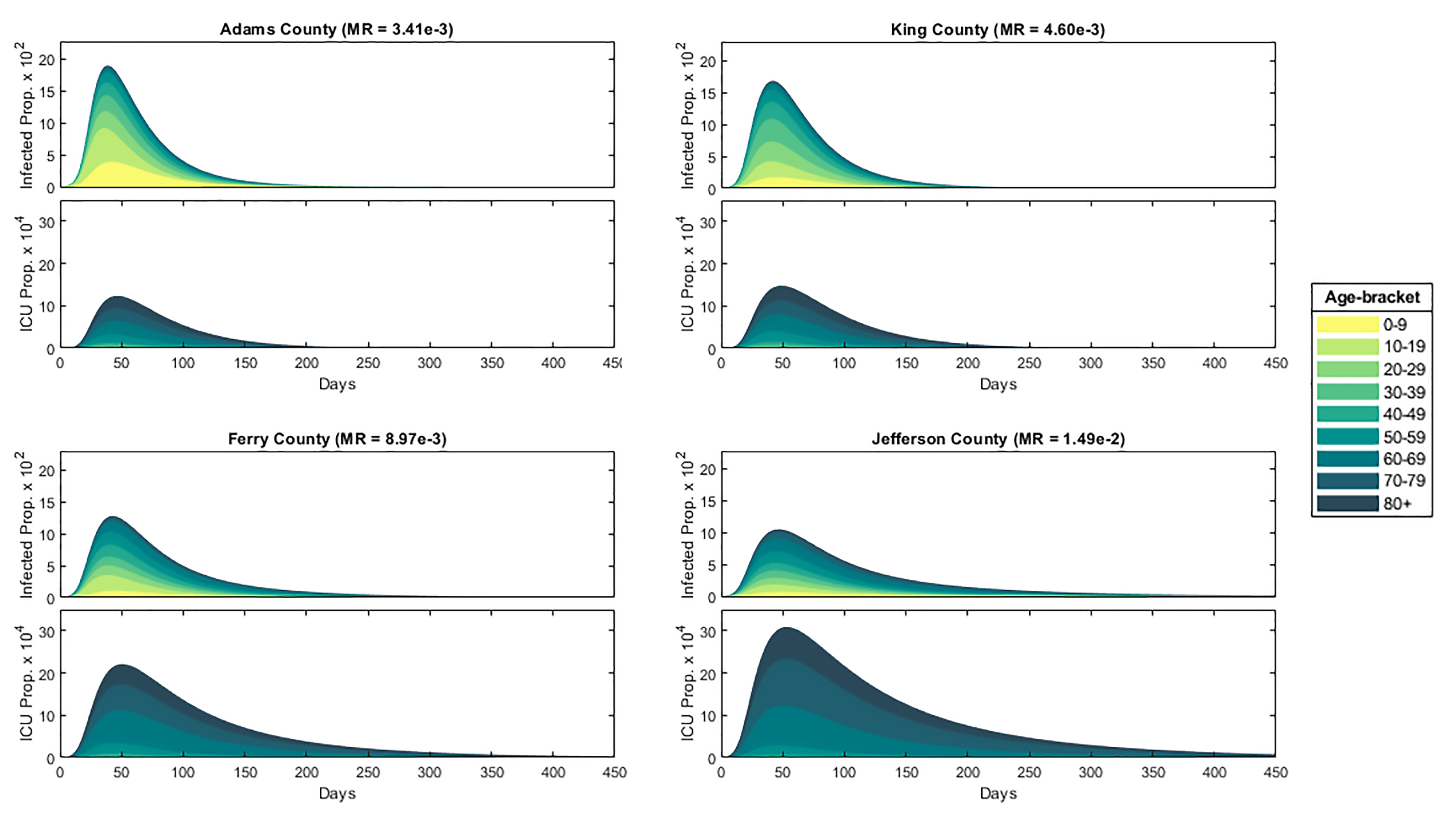}
 \caption{Homogeneous Mitigation Comparison, where the model was run until the proportion of the population infected fell below $10^{-6}$.}
\label{fig:hComp}
\end{figure}
     \end{center}

In order to understand the relation between the proportion of the population infected and  the median age of a county, we consider Figure \ref{fig:hComp} which displays the current proportion of the population infected and in intensive care over time for the four selected counties.   Mitigation measures are the same as shown in \ref{fig:lambda}, applied with $\lambda = 0.1$ across all age-brackets homogeneously.  
 In particular, we can see that the proportion of the population infected decreases with increasing median age. 
 This is likely due to a larger mitigation factor (stricter social-distancing policy) associated with the greater ICU population as well as a larger number of total contacts among the younger population from Figure \ref{fig:cmat1}.

 \onecolumngrid
 \twocolumngrid

%\begin{figure}[H]
%\centering
%\subfigure[\ Adams County]{\includegraphics[width=\linewidth]{AdamsH.png}}
%\subfigure[\ King County]{\includegraphics[width=\linewidth]{KingH.png}}
%\subfigure[\ Ferry County]{\includegraphics[width=\linewidth]{FerryH.png}}
%\subfigure[\ Jefferson County]{\includegraphics[width=\linewidth]{JeffersonH.png}}
%
%\caption{Homogeneous Mitigation Comparison}
%\label{fig:hComp}
%\end{figure}

\begin{figure}[H]
    \begin{center}
    \includegraphics[width = 0.95\linewidth]{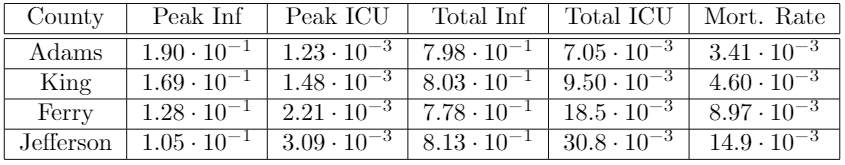}
    \end{center}
    \caption{County Infected/ICU Statistics}
    \label{fig:tableH}
\end{figure}

The {\it Herd Immunity Threshold} (HIT) is the critical proportion of the population that must become immune for an epidemic to longer persist. In an SIR model, the HIT value is given by $1-\frac{1}{R_0} \approx 82 \%$ for $R_0 = 5.7$.
 In order to understand the HIT for our model, we first consider the peak and total proportion infected and in intensive care for each of the four counties, as seen  in Figure \ref{fig:tableH}. 

Note that in all four counties, around the same proportion of the population became infected while a significant proportion (about $20\%$) never became infected throughout the course of the epidemic. Therefore a state of herd immunity was achieved in which a large enough proportion of the population achieved immunity though previous infections, thereby reducing the probability of new infections, eventually halting the spread of the disease. 

\begin{figure}[H]
    \begin{center}
    \includegraphics[width = 0.95\linewidth]{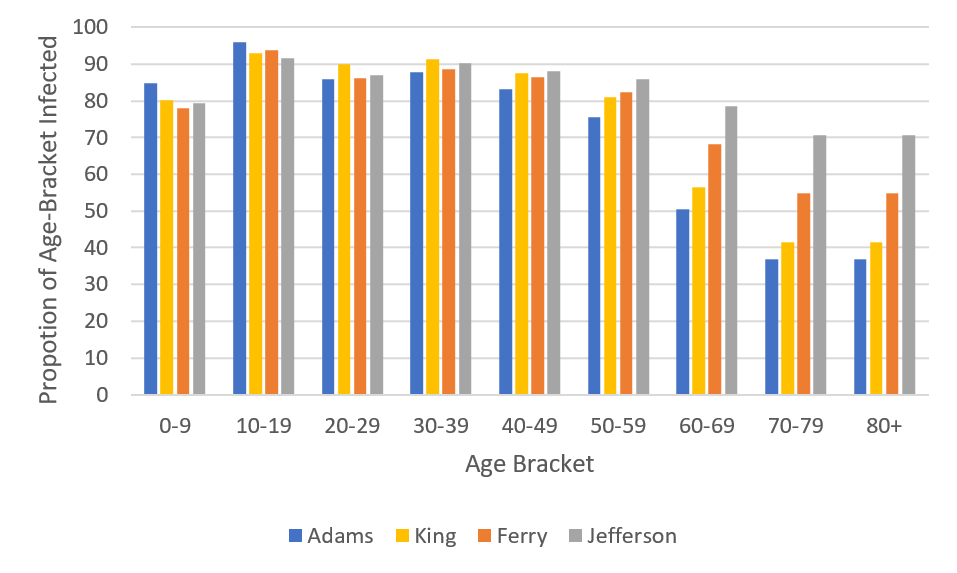}
    \end{center}
    \caption{Proportion of Age-Bracket Infected}
    \label{fig:infDistH}
\end{figure}

Since we are interested in understanding effects of mitigation strategies for  the less vulnerable population ($<60$ years),  we consider in Figure \ref{fig:infDistH}  the proportion of each individual age-bracket infected in the epidemic for each county, i.e. the probability that an individual will be infected given their age-bracket. Through our study, we find that for the less vulnerable population ($<60$ years), the probability of infection is roughly the same regardless of age-bracket and population age-distribution, as can be seen in in Figure \ref{fig:infDistH}.

On the other hand, for the more vulnerable population ($\ge 60$ years), the probability of infection increases significantly with the median age of the population. As a result, counties such as Ferry and Jefferson not only have a larger vulnerable population, but also have a larger proportion of their vulnerable population infected, which greatly contributes to their mortality rate.

\subsection{Effects of Age-Specific Policy}
By shifting the distribution of the infected population away from the vulnerable population, the mortality rate of an epidemic can be reduced significantly. In what follows  we shall examine the effect of age-specific policy including partial opening of schools and workplaces that prioritizes and targets the more vulnerable over less vulnerable populations.

We first examine the effects of relaxing school and work restrictions.  For each scenario, we choose a relaxed bracket: the part of the population unaffected by the social distancing policy (that scales the contact-matrix by the mitigation-factor). 

\begin{figure}[H]
    \begin{center}
    \includegraphics[width = 0.8\linewidth]{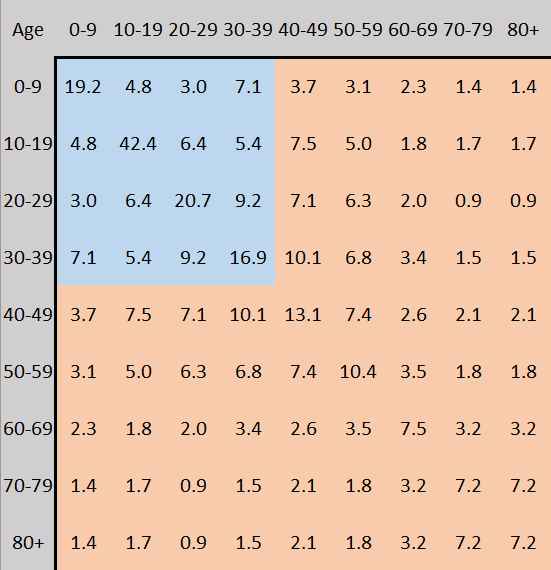}
    \end{center}
    \caption{Population affected in relaxing school restrictions}
    \label{fig:scaMat}
\end{figure}

For example, when relaxing school restrictions, the targeted population is all individuals $<30$ years, meaning all contacts amongst this group (the blue group in Figure \ref{fig:scaMat}) will not be subject to restrictions, while the remaining contacts (the red group) will be subject to normal restrictions given by the mitigation-factor. For relaxing work restrictions, a similar group is relaxed, targeting all individuals $<70$ years. Note that relaxing work restrictions is applied on top of relaxing school restrictions as the school age-bracket is a subset of the work age-bracket.

%\newpage
%\begin{figure}[H]
%    \begin{center}
%    \includegraphics[width = \linewidth]{chartPS.png}
%    \end{center}
%    \caption{School bracket relaxed ($<30$ years)}
%    \label{fig:PSchart}
%\end{figure}

%\begin{figure}[H]
%    \begin{center}
%    \includegraphics[width = \linewidth]{chartPW.png}
%    \end{center}
%    \caption{Work bracket relaxed ($<60$ years)}
%    \label{fig:PWchart}
%\end{figure}

Comparing the statistics from the table in Figure \ref{fig:tableP} below, to that from  the table in Figure \ref{fig:tableH} presented before, we find that, on average, peak infections increased by $58 \%$ when relaxing schools and increased $160 \%$ when relaxing work. In both cases, total infections increased slightly, continuing to remain around the herd immunity threshold of $82 \%$ as expected.\\

\begin{figure}[H]
    \begin{center}
    \includegraphics[width = \linewidth]{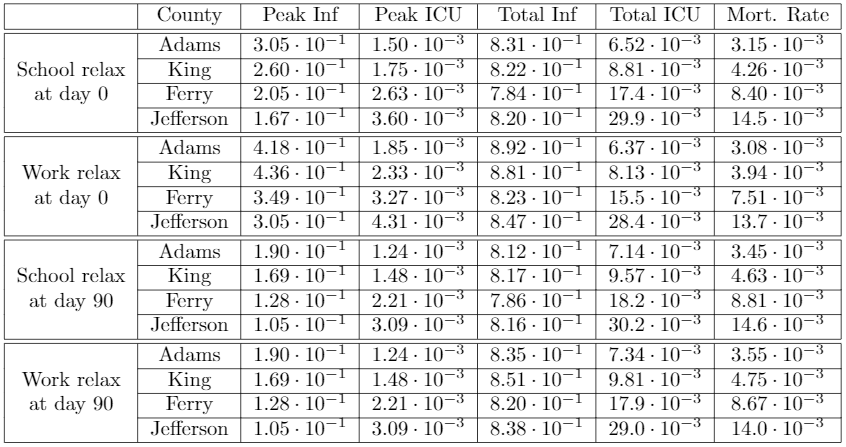}
    \end{center}
    \caption{Age-Specific Policy Statistics}
    \label{fig:tableP}
\end{figure}

\begin{center}

 \begin{figure}[H]
    \includegraphics[scale = 0.151]{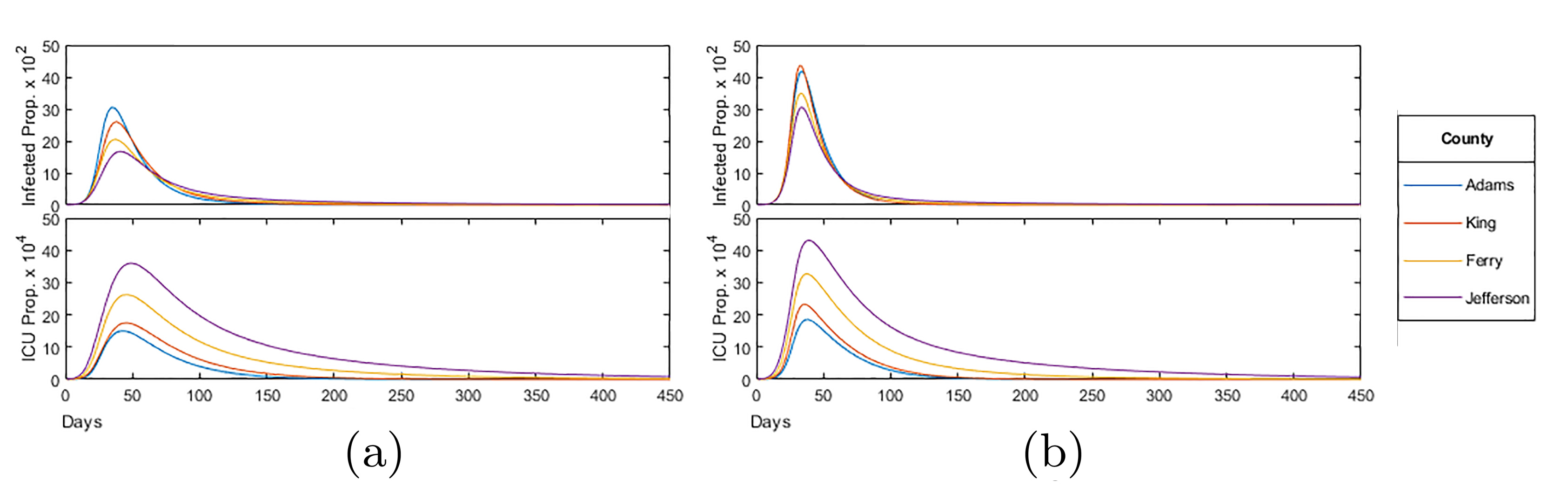}
 \caption{(a) School bracket relaxed ($<30$ years); (b) Work bracket relaxed ($<60$ years).}
\label{fig:PSchart}
\end{figure}
     \end{center}

  The current proportion of the population infected and in intensive care over time with fully relaxed school and work restrictions respectively can be seen in  Figures \ref{fig:PSchart} (a) and \ref{fig:PSchart} (b). As infections among the relaxed bracket increased drastically in proportion to the restricted bracket, the mean of the age-distribution of the infected population shifted towards the younger, less vulnerable, bracket by the time the HIT was achieved. As a result, we saw that the mortality rate, on average, decreased by $6.0 \%$ when relaxing schools and $12.1 \%$ when relaxing work, with this being less pronounced in the greatest median age Jefferson county with a $2.7 \%$ and $8.1 \%$ reduction respectively.

 \onecolumngrid
 \twocolumngrid

  \onecolumngrid
 \begin{center}

 \begin{figure}[h!]
    \includegraphics[scale = 0.153]{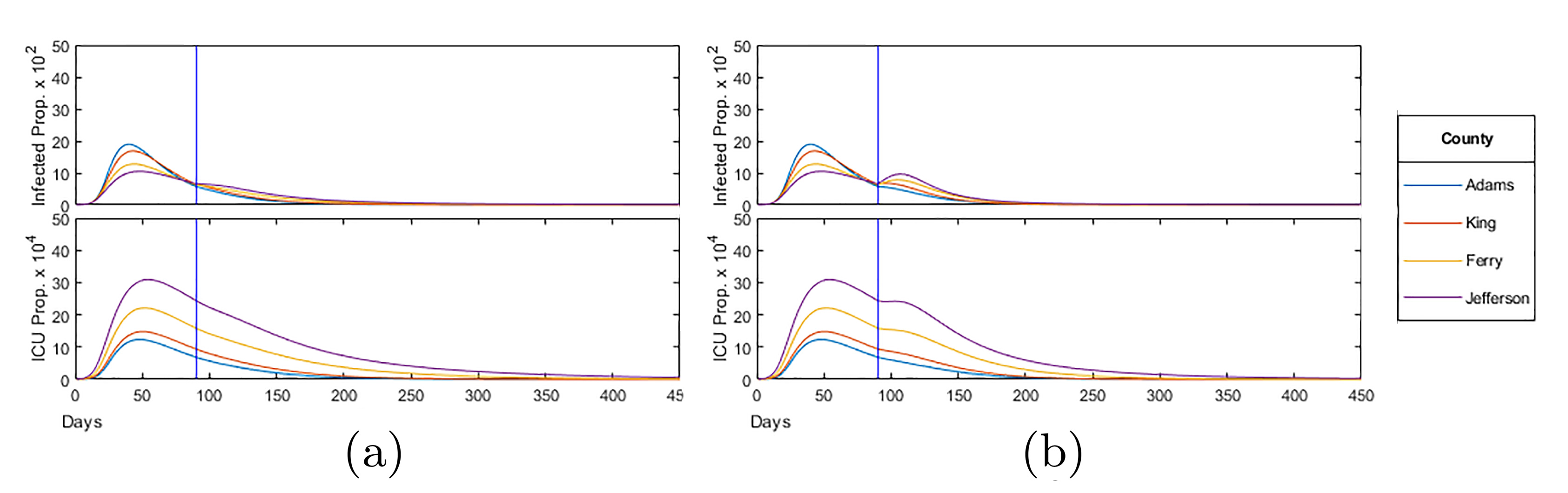}
 \caption{(a) School bracket relaxed ($<30$ years) at $90$ days; (b) Work bracket relaxed ($<60$ years) at $90$ days.}
\label{fig:PWLchart}
\end{figure}
     \end{center}
 \twocolumngrid

In contrast to the above, it should also be noted that, on average, peak ICU occupancy increased by $18.9 \%$ when relaxing schools and $51.2 \%$ when relaxing work, significantly more than the increase in peak infections. Reducing transmissions across the younger age-brackets has the effect of ``flattening the curve'': reducing peak infections and ICU occupancy while infecting roughly the same proportion of the population over a larger span of time. Although relaxing school and work restrictions reduced the calculated mortality rate, in practice, increasing peak ICU occupancy by up to $50 \%$ can put excessive strain on hospitals that are at full capacity, leading to additional moralities from a lack of resources needed to treat everyone requiring intensive care.

We also examine the effect of relaxing school and work restrictions after 90 days, roughly a month after the initial peak in infections. Figures \ref{fig:PWLchart} (a) and \ref{fig:PWLchart} (b) display the current proportion of the population infected and in intensive care over time with school and work restrictions relaxed at 90 days respectively. The vertical blue line indicates when the restrictions are relaxed for the targeted age-bracket. We find that relaxing school restrictions at 90 days (a) has little effect on the subsequent trajectory of the epidemic for all counties, with no change in peak infections and ICU, and mortality rate increasing by an average of $3.5 \%$ compared to constant restrictions in Figure  \ref{fig:tableH}. When relaxing work after 90 days (b), we see a notable change in the trajectory of the epidemic in counties with a higher median age. In Jefferson county, infections reached reached a new peak of $1.05 \cdot 10^{-1}$, identical to the peak before the relaxation. ICU occupancy in the county also increased for a period of 30 days following the relaxation. Among all counties, the total infected increased by an average of $3.7 \%$, exceeding the HIT for all counties, though the mortality rate remained mostly unaffected, decreasing by an average of $0.5 \%$.

%\begin{figure}[H]
%    \begin{center}
%    \includegraphics[width = \linewidth]{chartPSL.png}
%    \end{center}
%    \caption{School bracket relaxed ($<30$ years) at $90$ days}
%    \label{fig:PSLchart}
%\end{figure}

%\begin{figure}[H]
%    \begin{center}
%    \includegraphics[width = \linewidth]{chartPWL.png}
%    \end{center}
%    \caption{Work bracket relaxed ($<60$ years) at $90$ days}
%    \label{fig:PWLchart}
%\end{figure}

\subsection{Effects of Age-Targeted Vaccination}
Vaccinations play a critical role in mitigating the effects of an epidemic. By directly preventing the susceptible population from contracting the disease, it is possible to achieve herd immunity in less time and significantly reduce peak infections and mortality rate. An important aspect of vaccines is the strategy considered by a government or society in order to achieve the desired proportion of vaccinated population -- and much research has been done in this direction for long time know infections (e.g., for Zika and Hepatitis B, as discussed in \cite{strat2,strat3}) as well as for recent viral outbreaks such as COVID-19 (e.g, see for example \cite{strat1}).

In what follows, through our {\it modified age-structured SIR model}, we shall examine the effect of prioritizing certain age-groups in vaccine distribution versus a homogeneous distribution across all age-groups. 
We model vaccinations by directly transferring individuals from the susceptible and removed groups. In particular, if vector $\mu$ represents the number of individuals in each age-bracket vaccinated at each day, then we have the following:
    
    \begin{eqnarray}
    \frac{\mathrm dS_i}{\mathrm dt} &=& -\beta \cdot \frac{S_i}{N} \cdot \sum_{j=1}^n \mathcal M_{ij} \cdot I_j - \mu_i ; \\ 
    \frac{\mathrm dI_i}{\mathrm dt} &=& \beta \cdot \frac{S_i}{N} \cdot \sum_{j=1}^n \mathcal M_{ij} \cdot I_j - \gamma \cdot I_i ;\\ 
    \frac{\mathrm dR_i}{\mathrm dt} &=& \gamma \cdot I_i +\mu_i.
\end{eqnarray}

To select the distribution of vaccines, we define a weight vector $\omega$ that represents the priority of each age-bracket in our distribution. 

\pagebreak

  \onecolumngrid
 \begin{center}

 \begin{figure}[h!]
    \includegraphics[scale = 0.12]{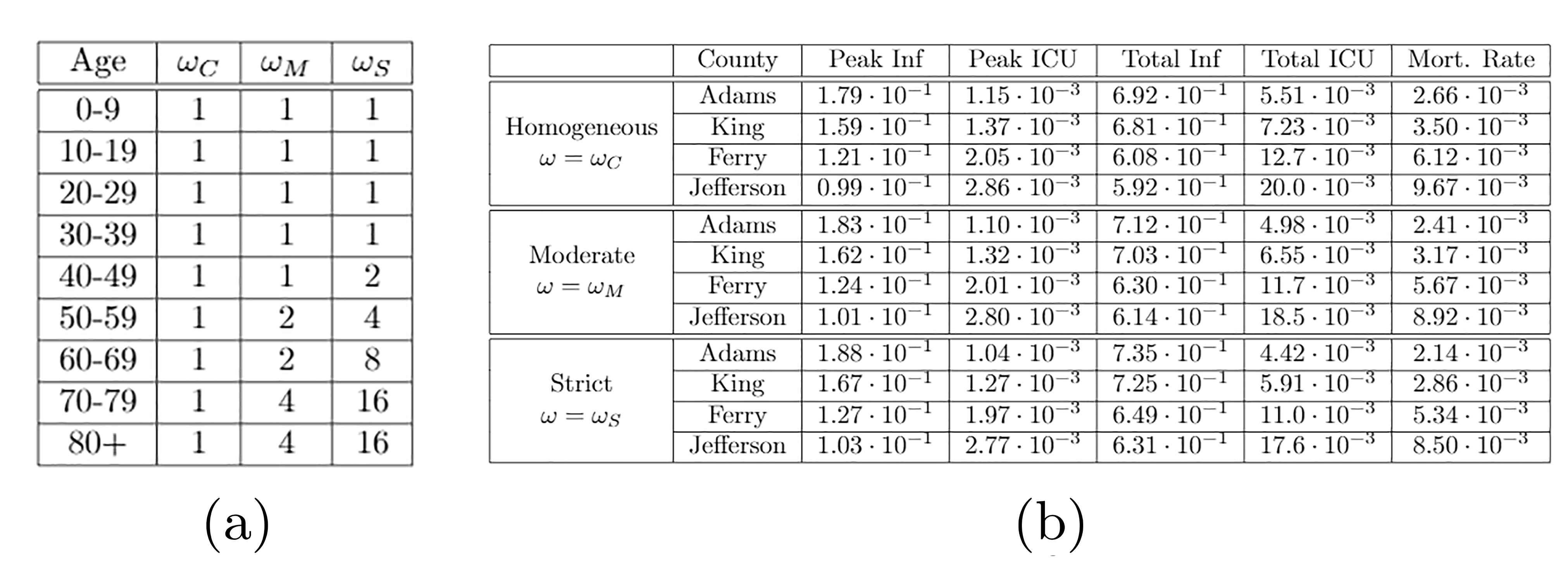}
 \caption{(a) Relative weights of vaccine distribution; (b) ge-Targeted Vaccine Distribution Statistics}
\label{fig:tableV}
\end{figure}
     \end{center}
 \twocolumngrid

Using the weight vector, we define \[\mu_i = T \cdot \frac{\omega_i \cdot S_i}{|\omega \odot S|},\] where $T$ is the total number of vaccines administered at each day and $|\omega \odot S|$ is the sum of the elements of vector $\omega \odot S$. Note that when $\omega$ is constant, each member of the susceptible population is equally likely to receive a vaccine. The United States produces enough flu vaccines yearly for approximately half of its population \cite{flu1}, and thus for our model we shall set $T=N/720$ where $N$ is the population size.  

For our age-targeted distributions we provide the higher age-brackets moderate priority with vector $\omega_M$ and strict priority with vector $\omega_S$.  We let $\omega_C$ denote the constant weight vector for the homogeneous (control) distribution. The weight values selected are summarized in the table of Figure \ref{fig:tableV} (a), and the model statistics for each distribution are given in Figure \ref{fig:tableV} (b).

% 
%\begin{figure}[H]
%\begin{center}
%    \includegraphics[width = 0.45\linewidth]{tableW.png}
%\end{center}
%\caption{Relative weights of vaccine distribution}
%\label{fig:chartW}
%\end{figure}
%\begin{figure}[H]
%    \begin{center}
%    \includegraphics[width = \linewidth]{tableV.PNG}
%    \end{center}
%    \caption{Age-Targeted Vaccine Distribution Statistics}
%    \label{fig:tableV}
%\end{figure}

\begin{center}

 \begin{figure}[h!]
    \includegraphics[scale = 0.155]{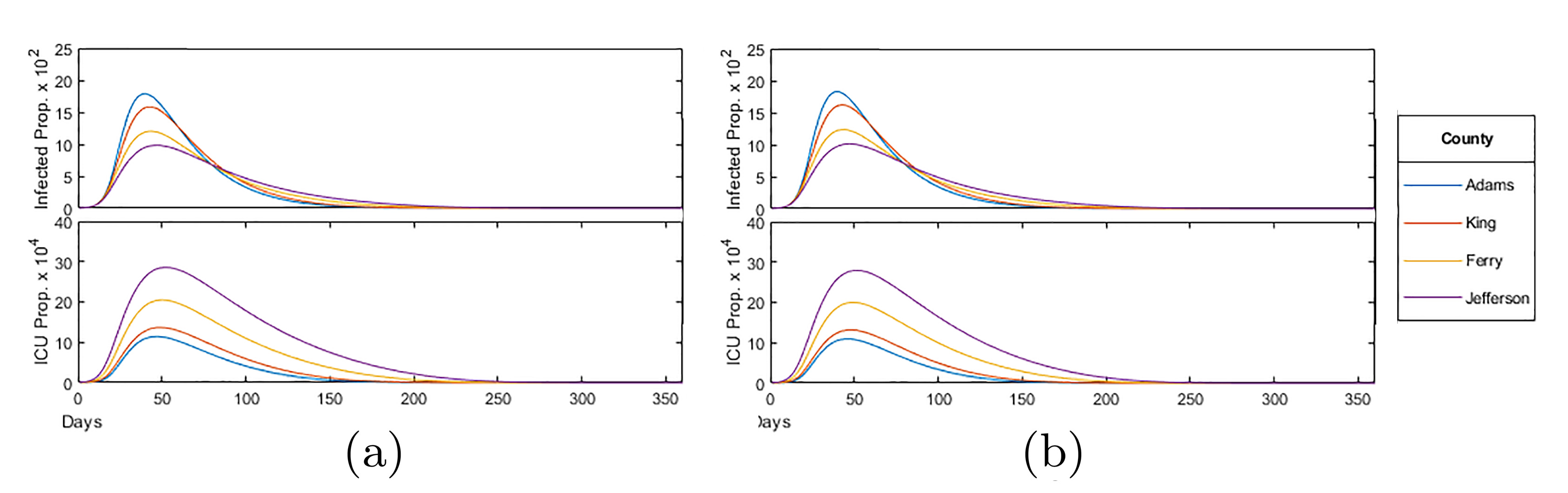}
 \caption{(a) Homogeneous vaccine distribution ($\omega = \omega_C$); (b) Moderate priority distribution ($\omega = \omega_M$)}
\label{fig:VaxHom}
\end{figure}
     \end{center}

We shall first consider the case of a homogeneous vaccine distribution $(\omega = \omega_C)$. In this case,   the infected count and ICU occupancy is shown in Figure \ref{fig:VaxHom} (a). 
%
%\begin{figure}[H]
%    \includegraphics[width = \linewidth]{chartHom.png}
% \caption{Homogeneous vaccine distribution ($\omega = \omega_C$)}
%\label{fig:VaxHom}
%\end{figure}
%
%
One can see, in particular, that the administration of vaccines had a significant mitigating effect on the epidemic, on average reducing:\begin{itemize}
\item the peak infections by $5.9 \%$, and
\item the peak ICU occupancy by $7.2 \%$,
\end{itemize}
 with the reduction being more prominent on higher median age counties. Additionally, 
 \begin{itemize}
 \item 
 the mortality rate was reduced by an average of $28.2 \%$;  
 \item the HIT was reduced, especially among high median age counties, with $69.2 \%$ of Ferry county and $59.2 \%$ of Jefferson county infected in total. 
 \end{itemize}
 
 This should be compared to the same analysis done for   the moderate  vaccine distribution $(\omega = \omega_M)$ shown in  Figures \ref{fig:VaxHom} (b).
  \onecolumngrid
 \twocolumngrid

One may also consider  a strict  priority vaccine distribution $(\omega = \omega_S)$, for which the analysis is shown in  Figures  \ref{fig:VaxStr} below.
\begin{figure}[H]
     \includegraphics[width = \linewidth]{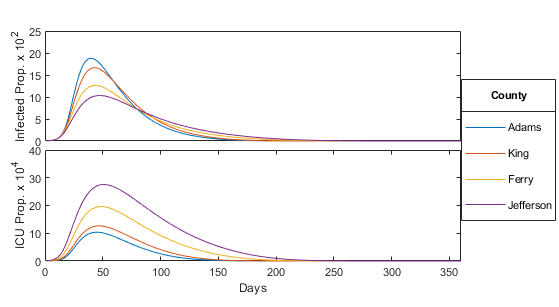}
 \caption{Strict priority distribution ($\omega = \omega_S$)}
\label{fig:VaxStr}
\end{figure}
 By increasing the relative weight of higher age-brackets, we present a model where we administer more vaccines towards the older and more vulnerable populations. As for general trends, we find that, compared to the $\omega_C$ distribution, increasing the priority of older age-brackets one has:
 
\begin{itemize}
\item an increases peak and and number of total infections, 
\item while a decreasing peak and total number of ICU occupancy. 
\end{itemize}
This trend is expected as prioritizing older age-brackets results in a greater proportion of the younger population susceptible to infection who are more likely to become infected and spread the virus, increasing total infections.

 Moreover, by reducing the proportion of the older population susceptible, we also reduce their infections and ICU occupancy, subsequently lowering the mitigation factor. Furthermore, the age-targeted distributions were highly effective in further reducing mortality rate, with an average $8.5 \%$ reduction for $\omega_M$ and $15.7 \%$ reduction for $\omega_S$. In particular, Adams, the low median age county, responded most effectively with a $9.4 \%$ and $19.5 \%$ reduction in mortality rate for $\omega_M$ and $\omega_S$ respectively. 

%\begin{figure}[H]
%    \includegraphics[width = \linewidth]{chartMod.png}
% \caption{Moderate priority distribution ($\omega = \omega_M$)}
%\label{fig:VaxMod}
%\end{figure}

\begin{center}

 \begin{figure}[h!]
    \includegraphics[scale = 0.15]{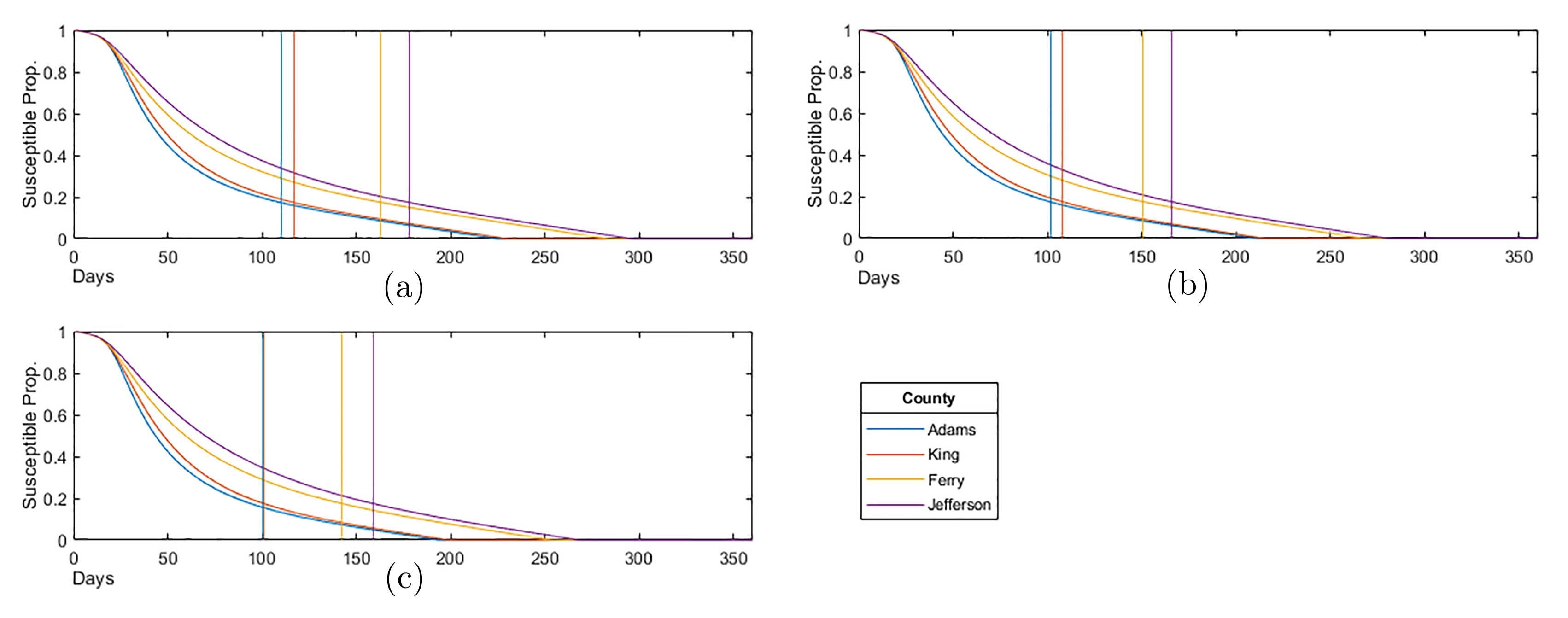}
 \caption{Susceptible Proportion for: (a) the homogeneous priority distribution ($\omega = \omega_C$); (b) the  moderate priority distribution ($\omega = \omega_M$); (c) the strict priority distribution ($\omega = \omega_S$)}
\label{fig:SusC}
\end{figure}
     \end{center}

To better understand how vaccinations limit the spread of the epidemic, we consider the proportion of the population susceptible to the virus over time in Figure \ref{fig:SusC}, %(a), (b), (c) for $\omega_C$, $\omega_M$, $\omega_S$ respectively. 
where the vertical lines indicate when each proportion susceptible in each corresponding county falls below (1 minus) the calculated HIT or $1/R_0 = 0.175$, summarized in Figure \ref{fig:tableHIT} below.\\

\begin{figure}[H]
    \begin{center}
    \includegraphics[width = 0.6\linewidth]{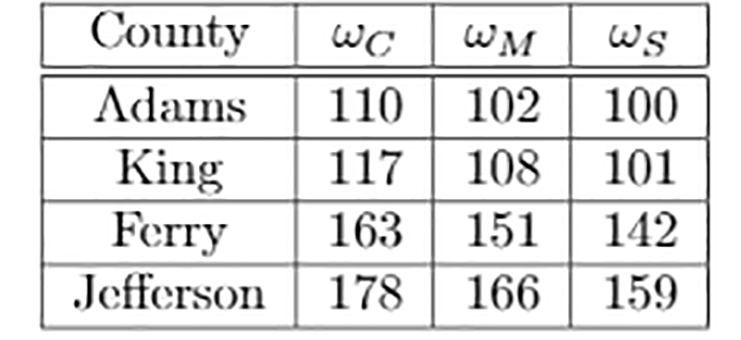}
    \end{center}
    
 \caption{Number of days until herd immunity is achieved}
\label{fig:tableHIT}
\end{figure}

 This is the point at which the spread of the virus no longer persists due to herd immunity, with the remainder of the susceptible population being immune through continued vaccinations. Although it results in greater total infections, we find that age-targeted vaccinations are effective in reducing the time required to achieve herd immunity, with an average $7.3 \%$ and $11.6 \%$ reduction in days for $\omega_M$ and $\omega_S$ respectively.

  \onecolumngrid
 \twocolumngrid

% 
%\begin{figure}[H]
%    \includegraphics[width = \linewidth]{new5.png}
% \caption{Susceptible Proportion for: (a) the homogeneous priority distribution ($\omega = \omega_C$); (b) the  moderate priority distribution ($\omega = \omega_M$); (c) the strict priority distribution ($\omega = \omega_S$)}
%\label{fig:SusC}
%\end{figure}
%
%\begin{figure}[H]
%    \includegraphics[width = \linewidth]{susM.png}
% \caption{Susceptible Proportion for moderate priority distribution ($\omega = \omega_M$)}
%\label{fig:SusM}
%\end{figure}
%
%\begin{figure}[H]
%    \includegraphics[width = \linewidth]{susS.png}
% \caption{Susceptible Proportion for strict prioirty distribution ($\omega = \omega_S$)}
%\label{fig:SusS}
%\end{figure}

\section{Conclusion and summary of our work}

In the present paper, we have introduced a modified age-structured compartmental SIR model using a function that scales contacts by a factor proportional to the current ICU occupancy in Washington State, USA, which serves to emulate the phased social distancing policy implemented Washington State.

 Our modeled epidemic utilizes the same disease parameters of the current COVID-19 pandemic with an $R_0$ value of $5.7$ and corresponding hospitalization, intensive-care, and mortality rates for each age-bracket, shown in Figure \ref{fig:agetable}. To understand the importance of age-targeted analysis of epidemic's contention, we apply our model across four populations: counties Adams, King, Ferry, and Jefferson of Washington State which possess varying population age distributions shown in Figure \ref{fig:ageDist} with medians 28.3, 36.8, 49.2, 57.2 respectively.

For our study, we first apply our unaltered model to each of the four counties, plotting the proportion of the population infected and in critical care over time in   Figure \ref{fig:hComp}. Through this, we find that
\begin{itemize}
\item[(i)] as the median age increases, peak infections decrease while peak ICU occupancy and mortality rate increases. 
\end{itemize}
This is due to younger age-brackets producing more contacts and infections over time, as seen in Figure \ref{fig:cmat1}, while the more vulnerable older age-brackets increasing ICU occupancy and placing heavier restrictions according to the rescaling factor. However, the total proportion of the population infected remain near the calculated herd immunity threshold ($1- \frac{1}{R_0} \approx 82 \%$) for all counties. As the herd immunity threshold is determined by disease rather than the population (assuming no external influences on the susceptible population such as through vaccination), we find that   the calculated mortality rate is mostly dependant on the age-distribution of the infected population.   

We then analyze the effect of age-specific policy such as fully relaxing restrictions on the school bracket (0-29 years) and the work bracket (0-69 years) both at the start of the epidemic (shown in Figure \ref{fig:PSchart}) and after the initial peak in infections at day 90 (shown in Figure \ref{fig:PWLchart}). Through our analysis, we found that 
\begin{itemize}
\item[(ii)] when relaxing school and work bracket at 0 days, peak infections increased on average by $58 \%$ and $160 \%$ respectively;
\item[(iii)]  while the proportion of the population infected remained around the HIT as expected.\end{itemize}
 As a result, relaxing restrictions across younger age-brackets lowered the median age of the infected population, leading to a decrease in mortality rate, with the reduction less pronounced in higher median age counties. However, we also saw that 
 \begin{itemize}
 \item[(iv)] peak ICU occupancy increased by an average of $18.9 \%$ when relaxing schools and $51.2 \%$ when relaxing work, significantly more than the increase in peak infections. 
 \end{itemize}
 Although the calculated mortality rate decreased, increasing peak ICU occupancy by up to $50 \%$ can overload the healthcare capacity in practice leading to additional, preventable deaths. Moreover, we saw that 
 \begin{itemize}
\item[(v)]  relaxing the school bracket after 90 days had little effect on the subsequent trajectory of the epidemic in all counties, with no change in peak infections and ICU occupancy;
\item[(vi)] relaxing the work bracket at the same time had notable effects on high median age counties.
\end{itemize}  In the particular case of  Jefferson, infections reached a new peak, identical to that before the relaxation and ICU occupancy also increased for a period of 30 days. However,  for all counties, the mortality rate remained mostly unaffected. 

Finally, we analyze the effect of age-targeted vaccine distribution. We model vaccinations by transferring a constant number of individuals from the susceptible to removed groups at each day. Under a normal homogeneous distribution ($\omega_C$), the number of individuals vaccinated in each age-bracket is proportional to the size of its susceptible population. In contrast,  in age-targeted distributions, we apply a set of weights as shown in Figure \ref{fig:tableV} (a),  so that individuals in certain age-brackets are more likely to become vaccinated, allowing us to target vaccinations towards more vulnerable age-brackets with moderate priority ($\omega_M$) and strict priority ($\omega_S$).

On its own, administrating vaccinations homogeneously as shown in Figure \ref{fig:VaxHom} had a significant mitigating effect on the epidemic with
\begin{itemize}\item[(vii)] an average $28.2 \%$ reduction in mortality rate compared to without vaccinations. 
\item[(viii)] the proportion of the population infected falling below the expected HIT, especially among Jefferson with the epidemic infecting only $59.2 \%$ of the population. 
\end{itemize}
When applying age-targeted vaccinations as shown in Figure \ref{fig:VaxHom} (b) and Figure \ref{fig:VaxStr}, we found that 
\begin{itemize}
\item[(iix)] peak infections slightly increased while peak ICU occupancy decreased.
\item[(ix)] a reducing mortality rate with an average $8.5 \%$ reduction for $\omega_M$ and $15.7 \%$ reduction for $\omega_S$ compared to the homogeneous distribution, with Adams responding most effectively. 

\end{itemize}
 This being due to a larger population of susceptible individuals in the younger age-brackets who are more likely to spread the infection while remaining less at risk for hospitalization. Finally, when plotting the proportion of the population susceptible, as shown in Figure \ref{fig:SusC},  we found that age-targeted vaccinations also 
 \begin{itemize}\item[(x)] reduce the time required for the epidemic to achieve the herd immunity threshold by an average $7.3 \%$ for $\omega_M$ and $11.6 \%$ for $\omega_S$.  
 \end{itemize}

\noindent{\bf Final Remarks.} 
To conclude our work, we shall present here different ways in which our model might be expanded,   as well as possible directions for future work. To provide better context for the extent of our result's implications, we list a series of key assumptions we have made, and which could be modified in order to make to expand on our model;:
\begin{itemize}
    \item Our mitigation coefficient to model social distancing policy is based strictly on a single parameter (ICU Occupancy). A more complex or modified mitigation coefficient may produce different results (e.g., one could consider economic factors, such as those studied in \cite{choi2020racial}). 
    
    \item In our SIR model, we do not consider asymptotic individuals and the possibility for re-infection (which in the case of COVID-19, one may want to consider \cite{gousseff2020clinical}). We have assumed that all individuals within an age-bracket are equally likely to become infected and transmit the disease. 
    \end{itemize}

Within our work we assume independence in the policy between different counties and do not consider the movement of individuals between populations, which is something that would be interesting to incorporate. 
Moreover, one should note that relaxing restrictions immediately affects all targeted age-brackets and has no affect on any contacts including individuals outside of these age-brackets. 
Finally, we have assume that vaccine production is constant throughout the course of the entire epidemic and we do not consider possible changes in supply and demand: it will be most interesting to incorporate the economic factors involved in vaccine production within an age-targeted study such as ours.

When considering other mitigation coefficients which could be used,  we see the following alternatives as potential paths for expanding our work further:

\begin{itemize}
    \item The infection rate ($\frac{\mathrm d}{\mathrm dt} I$) is another metric used to dictate policy in states such as New York and California, which could be considered. 
       Percentage positive tests, measured as the proportion of the population infected ($|I|/N$), is another factor in states such as North Carolina and Georgia used to indicate the extent of disease spread.

    \item Instead of gradual/proportional restrictions, isolated populations such as those of New Zealand, implemented strict lockdowns within the first cases with the goal of eradicating the disease before any possibility of herd immunity. Stricter policy can be modelled by increasing the $\lambda$ factor or scaling contacts by a factor of $|I|$ or $|I|^2$ within our model. 
    
\end{itemize}

 \noindent {\bf Disclaimer.} As with all mathematical models that are applied to real world systems, our results are  valid only under the model's assumptions. As such, the goal of our research is not to convey specific public health information and risks, but rather be a tool for health strategists for better planning and awareness with respect to policy. \\

 \noindent {\bf Acknowledgments.} The authors   are thankful to MIT
PRIMES-USA for the opportunity to conduct this research together, and  to James Unwin for inspiring conversations. The work of Laura P. Schaposnik is partially supported through the NSF grants  CAREER DMS 1749013, and she is thankful for the flexibility that the University of Illinois at Chicago, and the Mathematics, Statistics and Computer Science department in particular, has given her during these difficult childcare-free months of the pandemic. 
\smallbreak
 
\noindent {\bf Affiliations.}\\
  (a) Milton High School, Milton, GA 30004, USA. \\
  (b)  University of Illinois, Chicago, IL 60607, USA.

\bibliography{PRIMES2020}{}
\bibliographystyle{plain}

\end{document}